\documentclass[aps,prb,twocolumn,superscriptaddress]{revtex4}

\usepackage{chngcntr}
\usepackage{amsmath, amssymb}    
\usepackage{graphicx}  
\usepackage{natbib}  
\usepackage{epstopdf}
\usepackage{subfigure}
\usepackage{pstricks}
\usepackage{color}
\usepackage{mathrsfs}
\usepackage{fixmath}
\usepackage[hypertexnames=false]{hyperref}
\hypersetup{colorlinks=true}
\usepackage[all]{hypcap} 

\newcommand{\nn}{\nonumber \\}

\begin{document}

\title{ Strange superconductivity near an antiferromagnetic heavy fermion quantum critical point}

\author{Y. Y. Chang}
\email{cdshjtr.ep02g@nctu.edu.tw}
\affiliation{Department of Electrophysics, National Chiao-Tung University, Hsinchu, 300 Taiwan, R.O.C.}

\author{F. Hsu}
 \affiliation{Department of Physics, National Tsing-Hua University, Hsinchu, 300 Taiwan, R.O.C.}
 
\author{S. Kirchner}
 \affiliation{Zhejiang Institute of Modern Physics, Department of Physics, Zhejiang University, Hangzhou, P.R.C.}
 
\author{C. Y. Mou}
  \affiliation{Department of Physics, National Tsing-Hua University, Hsinchu, 300 Taiwan, R.O.C.}
  \affiliation{Physics Division, National Center for Theoretical Sciences, Hsinchu, 300 Taiwan, R.O.C.}
 
\author{T. K. Lee}
\affiliation{Institute of Physics, Academia Sinica, Nankang, Taipei, Taiwan, R.O.C.} 
 
\author{C. H. Chung}
\email{chung@mail.nctu.edu.tw}
\affiliation{Department of Electrophysics, National Chiao-Tung University, Hsinchu, 300 Taiwan, R.O.C.}
 \affiliation{Physics Division, National Center for Theoretical Sciences, Hsinchu, 300 Taiwan, R.O.C.}
 \affiliation{Institute of Physics, Academia Sinica, Nankang, Taipei, Taiwan, R.O.C.}

 
  \date{\today}
       
       
      \begin{abstract}
The heavy fermion Ce$M$In$_5$ family with $M$ = Co, Rh, Ir provide a prototypical example of strange superconductors with unconventional $d$-wave pairing and strange metal normal state, emerged near an antiferromagnetic quantum critical point. The microscopic origin of strange superconductor and its link to antiferromagnetic quantum criticality and strange metal state are still open issues. 
We propose a microscopic mechanism for strange superconductor,
 based on the coexistence and competition between the Kondo correlation and the quasi-$2d$ short-ranged antiferromagnetic resonating-valence-bond spin-liquid near the antiferromagnetic quantum critical point via a large-$N$ Kondo-Heisenberg model and renormalization group analysis beyond the mean-field level. We find the coexistence (competition) between the two types of correlations well explains the overall features of superconducting and strange metal state. The interplay of these two effects provides a qualitative understanding on how superconductivity emerges from the SM state and the observed  superconducting phase diagrams for Ce$M$In$_5$ near the AF QCP.
\end{abstract}       
       \maketitle
{\it Introduction.} Heavy fermion superconductivity (HFSc) has attracted much attention both theoretically and experimentally due to the incompatible nature of superconductivity and magnetism [\onlinecite{Steglich-CeCuSi}].
One particular interesting class of HFSc is that of ``strange superconductors" (SSc) where Cooper pairs get condensed out of a strange metal (SM) state of incoherent non-Fermi liquid (NFL) excitations reminiscent of what has been found in the cuprate superconductors [\onlinecite{Taillefer-HiTc}]. A large specific heat jump at $T_c$ [\onlinecite{Thompson-CeCoIn-SC, Thompson-PRL-2001}] and large effective mass seen in London penetration depth [\onlinecite{Matsuda-CeCoIn-lambda}] and superconducting coherence length suggest SSc involve pairing of the heavy $f$-electrons. The unconventional ($d$-wave) superconductors family CeCoIn$_5$ and CeRhIn$_5$ [\onlinecite{Thompson-d-wave, Davis-d-wave, Coleman2007}] offers a proptotypical example of this kind, displaying SM normal state with $T$-linear resistivity and power-law singularities in specific heat and magnetic susceptibility [\onlinecite{Thompson-CeCoIn-SC}].  The Kondo hybridization between local $4f$ and mobile $5d$ electrons of the Ce atoms [\onlinecite{CeCoIn-1st-principle, Feng-arpes-CeCoIn}] as well as the antiferromagnetic (AF) Ruderman-Kittel-Kasuya-Yosida (RKKY) coupling between the $f$-electrons [\onlinecite{Davis-STM-PNAS}] have been suggested to play important roles in superconductivity. A superconducting dome emerges near the antiferromagnetic quantum critical point (QCP) under pressure and magnetic field [\onlinecite{ThompsonPRL-2011}]. 
\begin{figure}[ht]
        	\centering
        	\includegraphics[scale=0.35]{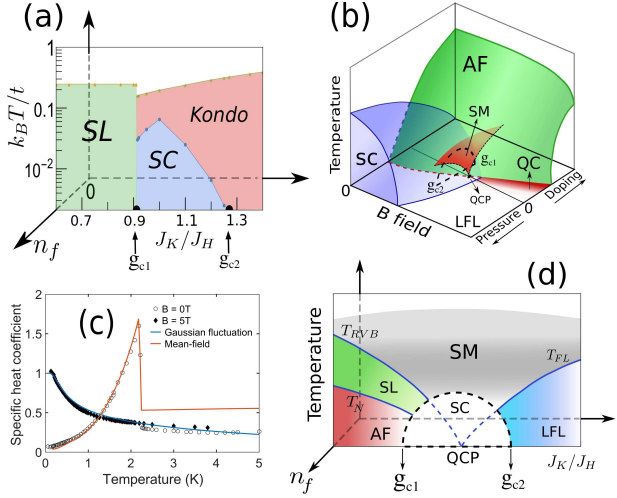}
        	\caption{(a) The mean-field phase diagram is generated self-consistently with the occupation number per site of the $f$-electron $n_f = 0.8$. (b) The ($p,~ B,~ T$) phase diagram of CeCoIn$_5$ is schematically adapted from Ref. \onlinecite{ThompsonPRL-2011}. The fan-shaped region in the middle represents the NFL SM region while the dome for the superconducting state at the mean-field level. The quantum critical (QC) line separates the AF and the LFL phases. (c) The red curve shows the (normalized) specific heat coefficient in the superconducting state at the mean-field level with $J_K/t = 1.2$, $\mu/t = -0.12$ and $J_H/t = 1.0$, fitted to the zero field data ($\bigcirc$). The blue curve shows the specific heat coefficient in the SM state, fitted to the data at $B = 5$T ($\blacklozenge$). (d) The proposed phase diagram near the QCP with $n_f \rightarrow 1^-$: A NFL strange metal (SM) state (grey region) emerges as a quantum critical region from the QCP, which separates the RVB spin-liquid phase (SL) (green region between $T_{RVB}$ and $T_N$) and the FL phase (blue area below $T_{FL}$). At lower temperatures, AF long-range order phase is expected to appear (red area below the N$\acute{\text{e}}$el temperature $T_{N}$).}
        	\label{fig:phase-diag}
        \end{figure} 

Though phenomenological approaches based on the two-fluid model [\onlinecite{Pine-2017-RPP}], spin fluctuation theory [\onlinecite{Davis-STM-PNAS}] as well as a Quantum Monte Carlo approach based on a two-impurity Kondo model featuring local quantum criticality [\onlinecite{QMSi-qmc}] were used to account for HFSc, a comprehensive microscopic mechanism is still needed. In addition, Anderson's resonating-valence-bond (RVB) theory of cuprate superconductivity [\onlinecite{Anderson-RVB}] was proposed to explain the HFSc due to similarities between them [\onlinecite{Coleman2007}]. Close to the edge of antiferromagnetism, the Kondo effect may not only stablize the RVB spin liquid (SL) against the magnetic long-ranged order by partially sharing the $f$-electron spins [\onlinecite{Coleman-stablized-SL, Coleman-SpN}], but also introduce hoping of the RVB bonds to the conduction band to form charged Cooper pairs, leading to a Kondo-RVB coexisting heavy electron superconductor. Along this line, a controlled large-$N$ generalization of the physical SU(2) particle-hole (p-h) symmetric Kondo-Heisenberg model on a 2$d$ square lattice based on the Sp($N$) group symmetry was proposed [\onlinecite{Colema-SpN, Coleman-SpN}], where a direct transition from a Curie paramagnet to HFSc near $T_c$ is observed, indicating that the pairing mechanism occurs simultaneously with the Kondo effect. 

However, recent ARPES and STM investigations of CeCoIn$_5$ [\onlinecite{Feng-arpes-CeCoIn, Maple-CeCoIn-arxiv}] suggest a rather high onset Kondo temperature $T^\ast_{\text{onset}} \sim 100\, \text{K}-200 \, \text{K}$ [\onlinecite{footnote-Tk-1-ion,La-doped-CeCoIn}], well above the Kondo lattice coherence temperature $T^\ast_{coh} \sim 50\,\text{K}$ and $T_c \sim 2.3\,\text{K}$. Also, resistivity measurement for this compound shows a rather narrow superconducting region [\onlinecite{Thompson-CeCoIn-SC, ThompsonPRL-2011}].
 More importantly, outstanding puzzles still remain unsolved: How does superconductivity emerge from the SM normal state? What are the links between the SM, Kondo coherence, superconductivity, and QCP? What is the mechanism of superconductivity coexisting with antiferromagnetism [\onlinecite{Thompson-CeRhIn}]? 

As a step towards addressing these questions, we propose in this letter a microscopic mechanism based on the competition and coexistence between Kondo and AF RVB correlations near QCP via a p-h asymmetric class of Sp($N$) Kondo-Heisenberg model.
 Our mean-field results explain remarkably well the narrow superconducting dome as well as various temperature scales ($T_c$, $T^\ast_{\text{onset}}$ and $T^\ast_{coh}$) in terms of coexisting Kondo and RVB correlations. More importantly, we construct an effective field theory beyond mean-field near the QCP. Via renormalization group (RG) analysis, we show that the competition between the critical Kondo and RVB fluctuations lead to the observed SM behavior in the normal state. The interplay between coexistence and competition of these two effects provide a qualitative understanding on the emergence of HFSc from SM and the observed superconducting phase diagrams near the AF QCP.

 {\it The Large-$N$ Kondo-Heisenberg Hamiltonian.} The full action of the model will be discussed in Eq. (\ref{eq:fulllagrangian}) below. For clarity's sake, we start from the large-$N$ (Sp($N$)) p-h asymmetric Kondo-Heisenberg Hamiltonian at the mean-field level: $H_{MF} = H_0 + H_\lambda + H_K + H_J$, where $H_{0}=  \sum_{\langle i, j \rangle; \sigma}\left[ t_{ij}c^{\dagger}_{i\sigma}c_{j\sigma}+h.c. \right] - \sum_{i\sigma} \, \mu \, c_{i\sigma}^{\dagger} c_{i\sigma}$ describes hopping of conduction ($c_{i\sigma}$) electrons, $H_{\lambda}  =  \sum_{i, \sigma}\lambda \left [f_{i\sigma}^{\dagger}f_{i \sigma}-2S \right]$ describes the local $4f$-derived ($f_{i\sigma}$) electrons with $\lambda$ being the Lagrange multiplier to impose the local constraint 
$\langle \sum_\sigma f_{i\sigma}^{\dagger}f_{i \sigma} \rangle = N\kappa$   
 where a constant $\kappa \equiv 2S$ 
ensures the fully screened Kondo effect [\onlinecite{ColemanSB}] and $\kappa <1$ leads to p-h asymmetry and valence fluctuations. 
     The antiferromagnetic RKKY interaction reads: $H_{J} =  \sum_{\langle i,j \rangle}J_{ij}{\bf S}_{i}^{f} \cdot {\bf S}_{j}^{f}
      =  \sum_{\langle i,j \rangle;\alpha, \beta } \left[ \Phi_{ij}\mathcal{J}^{\alpha \beta} f_{i\alpha}f_{j\beta}+h.c. \right]+ \sum_{\langle i,j \rangle}N \frac{|\Phi_{ij}|^{2}}{J_{H}}$ is described in terms of fermionic Sp($N$) fields and an exchange coupling matrix with $\mathcal{J}^{\alpha \beta}=\mathcal{J}_{ \alpha\beta}=-\mathcal{J}^{\beta\alpha }$, note that Sp(1) $\sim$ SU(2) [\onlinecite{ReadSachdev1991, SachdevKagome, Chung2001}].
   The Kondo interaction is given by $H_{K} =  J_{K} \sum_{i} {\bf S}_{i}^{f} \cdot {\bf s}^{c} 
=  \sum_{i,~\sigma} 
\left[ \left( c_{i \sigma}^{\dagger} f_{i \sigma}  \right) \chi_{i} +h.c. \right] + \sum_{i}\, N\frac{|\chi_{i} |^{2}}{J_K}$. 
We assume a uniform RKKY coupling $J_{ij}=J_H$ on a lattice where $i,~j$ are nearest-neighbor site indices, $\sigma,\alpha,\beta \in \lbrace -\frac{N}{2}, \cdots, \frac{N}{2} \rbrace$. The mean-field Kondo hybridization and RVB spin-singlet bond are defined as $ \chi_{i} \equiv \langle \frac{J_K}{N}\sum_{\sigma} f^{\dagger}_{i\sigma}c_{\sigma}\rangle$ and $ \Phi_{ij} \equiv \langle \frac{J_{H}}{N} \sum_{\alpha, \beta} \, \mathcal{J}_{\alpha \beta}f^{\alpha \dagger}_{i}f^{\beta \dagger}_{j}\rangle$, respectively. 

{\it Coexisting Kondo-RVB Induced Superconductivity.} By varying the ratio $g \equiv J_K/J_H$, three distinct mean-field phases are found (see Fig. \ref{fig:phase-diag}(a)). Since with increasing pressure the antiferromagnetic state of the system is suppressed while the heavy Fermi liquid state is favored [\onlinecite{ThompsonPRL-2011}], we expect $g$ plays a similar role as pressure. In the RKKY limit ($g < g_{c1}$) and for $\kappa <1$ (equivalent to the average spin per site $S < 1/2$ due to valence fluctuation [\onlinecite{Coleman-nf-PRL, Fisk-nf-xray-PRB}]), a metallic spin-liquid (SL) phase (green area in Fig. \ref{fig:phase-diag}(a)) is found with short-range antiferromagnetic order parameter $\Phi_{ij}>0$ and $\chi_i = 0$. 
This phase is known as the fractionalized Fermi-liquid realized in frustrated Kondo lattice system [\onlinecite{Senthil-PRL-2003}]. In the opposite limit ($g > g_{c2}$), the Kondo state (red area in Fig. \ref{fig:phase-diag}(a)) emerges with $|\chi_i| > 0$ and $\Phi_{ij} = 0$. In the intermediate regime,  $g_{c1}<g<g_{c2}$, we find a Kondo-RVB coexisting superconducting (SC) phase (blue area in Fig. \ref{fig:phase-diag}(a)) $|\chi_i|>0$ and $|\Phi_{ij}|>0$ with a superconducting gap defined as $\Delta = \Delta (\textbf{k}) \equiv \Phi (\textbf{k})$  ($\Phi (\textbf{k})$ being the Fourier transformed $\Phi_{ij}$ with $d$-wave symmetry and $T_c$ being determined via $\Delta (T=T_c) = 0$, see Section A of Ref. \onlinecite{supp}). 
The superconductivity in the coexisting phase arises through higher-order processes involving both RVB and Kondo terms: $H_{sc} = \sum_{\langle i, j\rangle} (\chi_i\chi_j\Phi_{ij}\mathcal{J}^{\alpha \beta} c_{i\alpha}c_{j\beta}+h.c.)$ [\onlinecite{Nevidomsky-LT}]. The superconducting pairing gap is generated from the spin-singlet RVB bonds, while the Kondo hybridization provides hopping of the $f$-electron RVB singlets to the conduction($c$) electron band and fosters their Cooper pairing [\onlinecite{Nevidomsky-LT}]. Valence fluctuations ($n_f<1$) are expected to favor the single $f$-electron hopping via Kondo hybridization and frustrate the RVB singlet bonds, leading to a finite range in $J_K/J_H$ of superconductivity. When the fluctuations beyond mean-field are included, the short-range RVB SL state is expected to give way to long-range antiferromagnetism below $T_N$ (see Fig. 1(d)). However, the Sp($N$) representation of our model is not appropriate to describe symmetry-broken magnetic long-range ordered phases. We therefore refrain from discussing the microscopic origin of AF order. Nevertheless, the tendency toward  antiferromagnetic ordering when superconducting fluctuations are included will be discussed below within the RG framework.
 
Our mean-field results show qualitative and to some degree quantitative agreement with what is seen in experiment.  Various characteristic temperature scales are estimated: $T^\ast_{\text{onset}} \sim 90 \text{K}$ (with $|\chi(T<T^\ast_{\text{onset}})|>0$) [\onlinecite{foot-T-onset, Coleman-book, Satoh-Tk, HewsonBook}], $T^\ast_{coh}\sim 75 K$ [\onlinecite{Georges-PRL-Tscales}], and $T_c \sim   2  \, \text{K}$. They show $T_c \ll T^\ast_{coh} \ll T_{\text{onset}}^\ast$ and $T^\ast_{coh}/T_c \sim 30$, all in reasonably good agreement with experiments; although no crystal electric field are taken into account. [\onlinecite{Thompson-CeCoIn-SC, Feng-arpes-CeCoIn}] (see Fig. A.1 in Section A of Ref. \onlinecite{supp}). 
The estimated dimensionless Kondo coupling $J_K \rho_0 = J_H/J_K \sim 0.8$ for CeCoIn$_5$ at zero field agrees reasonably well with the measurement: $J_K \rho_0 = - \left(\ln \left(k_B T^\ast_{\text{onset}}/ D \right)\right)^{-1} \approx 0.56 - 0.91$
, where $T^\ast_{\text{onset}}$ shows a relation similar to the single-impurity Kondo scale [\onlinecite{Georges-PRL-Tscales}] 
with $\rho_0$ being the density of states at Fermi level (see Fig. A.2 in Section A of Ref. \onlinecite{supp}).
 Moreover, the (normalized) specific heat coefficient $C_V/T$ in the superconducting state is also well reproduced  [\onlinecite{Tinkham}] (see the red curve in Fig. \ref{fig:phase-diag}(c)), including a large jump at $T_c$ and a linear-in-$T$ dependence at low $T$ due to the $d$-wave nodal gap (see Section A of of Ref. \onlinecite{supp}). 
 These agreements support the idea that HFSc in the 115 family is mediated by the local valence-fluctuating $f$-electrons via coexisting RVB and Kondo hybridization, and also serve as a realistic basis for our analysis beyond the mean-field level.
        {\it The Strange Superconductivity and Quantum Criticality.} The key issue here is to  is to address the link between superconductivity, SM region, and the AF QCP. Experimental evidences on Gr\"{u}neisen parameter and thermal expansion coefficient [\onlinecite{ThompsonPRL-2011}] suggest that this QCP is located at the border of the superconducting, the antiferromagnetic and the Landau Fermi liquid (LFL) phases, is approached when superconductivity is suppressed (experimentally by the magnetic field and Cd doping (or negative pressure), see the quantum critical (QC) line in Fig. \ref{fig:phase-diag}(b) [\onlinecite{foot-nf-p9, Thompson-Cd-CeCoIn5}]). The SM region in Fig. \ref{fig:phase-diag}(d), proposed to be the quantum critical region associated with QCP, has been experimentally observed [\onlinecite{Thompson-CeCoIn-SC}]. Near this QCP our mean-field results show almost vanishing superconducting phase with nearly decoupled Kondo and the RVB phases (see Figs. \ref{fig:phase-diag}(d) and A.3 of Section A.1 of Ref. \onlinecite{supp}). 
        
         To go beyond mean field, we introduce the bosonic Gaussian fluctuation of the Kondo hybridization (RVB spin-singlet bond) via $\chi_i (\Phi_{ij}) \rightarrow \chi_i (\Phi_{ij}) +  J_\chi\hat{\tilde{\chi}}_i \, (J_\Phi \hat{\tilde{\Phi}}_{ij})$. An effective action beyond the Sp($N$) mean-field at $N=1$ near $g_{c1}$ and $g_{c2}$ in Fig. \ref{fig:phase-diag}(b), $S_{eff} = S_0 + S_{\chi} + S_\Phi +  S_G  + S_K +S_J +S_{sc}$, in the quasi-$2d$ limit $d=2+\epsilon$ with $0 <\epsilon \ll 1$ [\onlinecite{Thompson-CeCoIn-SC, foot-quasi-2d}] and the dynamical exponent $z = 2$ [\onlinecite{CHC-strangemetal}] is given by:
      \begin{widetext}
      \begin{eqnarray}
             S_{0}&= & \int dk \sum_{\sigma = \uparrow \downarrow}\, \  c_{k \sigma}^{\dagger} \, \left( -i \omega +\varepsilon_{c}({\bf k}) \right) c_{k\sigma}+f^{\dagger}_{k\sigma} \, \left( - \frac{i \omega}{\Gamma} + \lambda \right) f_{k\sigma}~,~ S_{K} =  J_{\chi}\sum_{\sigma=\uparrow\downarrow}\int dk dk^{\prime} \left[(c^{\dagger}_{k\sigma}f_{k^{\prime}\sigma})\hat{\tilde{\chi}}_{k+k^{\prime}}^{\dagger} +h.c.\right] ~, \nn
              S_{\chi} &=& \int \, dk \sum_{\sigma = \uparrow \downarrow} \left[\chi_{\textbf{k}}  f^{\dagger}_{k \sigma} c_{k\sigma} + h.c. \right] + \sum_{i} \int\,d\tau  |\chi_{i}|^{2}/J_{K}~,~
          S_{\Phi} = \int \, dk \sum_{\alpha \beta } \left[\Phi_{\textbf{k}} \epsilon_{\alpha\beta} f^{\alpha}_{k }f^{\beta}_{-k}  + h.c. \right] + \sum_{\langle i,j \rangle} \int \, d\tau |\Phi_{ij}|^{2}/J_{H} ~,\nn
 S_{J} &=& J_{\Phi} \sum_{\alpha,\beta =  \uparrow \downarrow } \int dk dk^{\prime} \left[\epsilon_{\alpha \beta}\hat{\tilde{\Phi}}_{k} f_{k^{\prime}}^{ \beta}f_{k+k^{\prime}}^{\alpha}+h.c.\right]~, ~ S_G =\int \, dk \left[\hat{\tilde{\chi}}^{\dagger}_{k}\, \left( -i \omega +\varepsilon_{\chi}({\bf k})+m_{\chi} \right) \hat{\tilde{\chi}}_{k}+ \hat{\tilde{\Phi}}^{\dagger}_{k}\, \left( -i \omega + \varepsilon_{\Phi}({\bf k})+ m_{\Phi} \right) \hat{\tilde{\Phi}}_{k}\right]~, \nn[5pt]
S_{sc} & =& \begin{cases}
 v_{sc} \sum_{(\alpha,\beta) = \uparrow,\downarrow}
\int dk_{1} dk_{2}dk_{3} \,  
  \left[ \chi_{{ k}_{1}}^{\dagger}\chi_{{ k}_{2}}^{\dagger}
\epsilon^{\alpha \beta}c_{{ k}_{3}, \alpha}^{\dagger}
c_{-{ k}_{1}-{ k}_{2}-{ k}_{3}, \beta}^{\dagger}+h.c.\right]\quad  (\text{near }g_{c1}) ~,\\[8pt]
						v_{sc}\sum_{(\alpha,\beta)=\uparrow\downarrow}\int d^{d}k \, d^{d}k^{\prime} \left[ \hat{\tilde{\Phi}}_{k} \epsilon^{\alpha \beta} c^{\dagger}_{  \alpha,k^{\prime}}c^{ \dagger}_{\beta,k+k^{\prime}} + h.c. \right] \quad (\text{near }g_{c2}),
\end{cases}
\label{eq:fulllagrangian}
\end{eqnarray} 
\end{widetext}
\begin{figure*}[ht]
              \centering
                \includegraphics[scale=0.15]{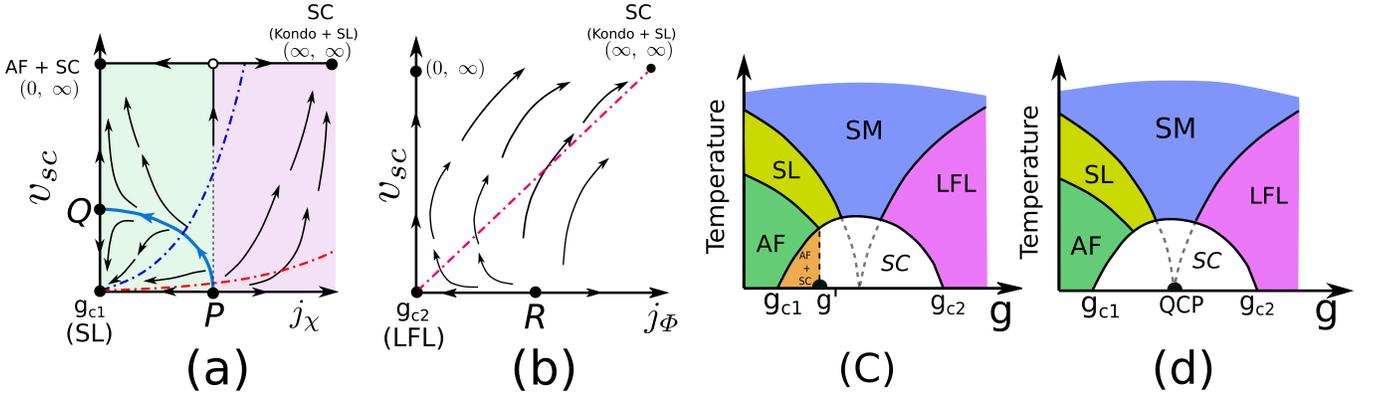}
                \caption{(a) and (b) show the RG flow diagrams near $g_{c1}$ and $g_{c2}$. In (a), both $j_\chi$ and $v_{sc}$ couplings flow to the Gaussian fixed point $g_{c1}$ for the initial couplings inside the blue solid curve connected $P$ and $Q$. 
                (c) and (d) are the schematic phase diagrams of CeRhIn$_5$ and CeCoIn$_5$, respectively. They can be qualitatively explained within our RG analysis. 
                }
                \label{fig:rgflow}
           \end{figure*} 
  where the $S_{sc}$ term arises from $H_{sc}$ as shown above. The competition of various couplings in $S_{eff}$ are investigated. In Eq. (\ref{eq:fulllagrangian}), $k\equiv (\omega, {\bf k})$ and $dk = d^{d}{\bf k}d\omega/(2\pi)^d$ and $\tau$ is the imaginary time. $\varepsilon_{c (\chi, \,\Phi)} (\textbf{k})$ is the quadratic dispersion for the conduction $c$-electrons (the fluctuation of the Kondo hybridization fluctuation $\hat{\tilde{\chi}}$, the fluctuation of the RVB spin-singlet bonds)(see Section B of Ref. \onlinecite{supp} and the Ref. \onlinecite{CHC-strangemetal} for details). Here, $m_{\chi (\Phi)} = 1/J_{K (H)}$ represents the bare mass of the $\hat{\tilde{\chi}}$ ($\hat{\tilde{\Phi}}$) bosons. We also introduce the factor $\Gamma$  to guarantee the local nature of the $f$-electrons. The actions $S_{\chi}$ ($S_K$) and $S_{\Phi}$ ($S_J$) represent the Kondo hybridization and RKKY interaction at (beyond) the mean-field level, respectively, and $S_G$ represents the action of the quadratic Gaussian fluctuating fields. The coefficient $J_{\chi (\Phi)}$ denotes the coupling constant for the Kondo (RKKY) term $S_{K (J)}$, whose bare value is taken to be $J_{K (H)}$. The term $S_{sc}$ is the action of the superconducting fluctuation from $H_{sc}$ defined above, an interplay between the AF RVB and Kondo correlations near the $g_{c1}$ and $g_{c2}$.
  
  When the superconductivity ($S_{sc}$) is suppressed, Eq. (\ref{eq:fulllagrangian}) has been shown to describe the NFL SM state as a quantum critical region separating RVB spin-liquid and Kondo states [\onlinecite{CHC-strangemetal}], which is relevant for the SM behavior seen in both pure and Ge-substituted YbRh$_2$Si$_2$ [\onlinecite{Taillefer-field-ind-115}]. It exhibits a linear-in-$T$ resistivity, a $T$-sublinear power-law divergence in magnetic susceptibility $\chi (T)\sim T^{-0.3}$ (see Section C of of Ref. \onlinecite{supp}), and a $T$-sublinear power-law divergent at low $T$ followed by a logarithmic-in-$T$ at higher $T$ in the electronic specific heat coefficient: $\gamma(T)  = \frac{C_V}{T} \sim |g-g_c|^{-\alpha} \, \Gamma \left(\frac{T}{T_{LFL}} \right)$ with the scaling function $\Gamma(x)$  behaving as $\Gamma (x) \sim x^{-\bar{\alpha} }$ for $x \simeq 1$ with $\bar{\alpha}$ being obtained by numerical fit of $\Gamma (x)$ and $-\Gamma (x) \sim \ln (x)$ for $1< x < 10$ (see the blue curve in Fig. \ref{fig:phase-diag}(c)) and $\alpha = \epsilon^2 + 3\epsilon/2$ being obtained via the scaling ansatz of free energy (see Ref. \onlinecite{CHC-strangemetal}). Both the observed anomalous $\chi (T)$ and $\gamma (T)$ are well accounted for by an extra spatial dimension being $\epsilon \sim 0.35$. In the crossover region $T_c < T < T^\ast_{coh}$, the resistivity behaves as $\rho = \rho_0 + A  T + B  T^2$ with $A$ and $B\ll A$ being prefactors, see Section D of of Ref. \onlinecite{supp}.
     
      With superconducting fluctuations ($S_{sc}$) included, we address how superconductivity emerges (disappears) around the two phase boundaries $g_{c1} \, (g_{c2})$ (shown in Fig. \ref{fig:phase-diag}(d)) using a diagrammatic renormalization group (RG) approach. We find that the key mechanism is the interplay between coexistence and competition of the Kondo and RVB interaction. Near $g_{c1}$, which separates the spin liquid from the superconducting phases, the ground state is the RVB spin-liquid phase ($\Phi \neq 0$ and $\chi = 0 $). As a result, the leading superconducting fluctuations are dominated by the fluctuations of Kondo hybridization $\hat{\tilde{\chi}}$ field with the RVB correlation being treated at the mean-field level i.e. $\hat{\tilde{\Phi}} \rightarrow 0$, and thus is described by $S_{sc}$ near $g_{c1}$ in Eq. (\ref{eq:fulllagrangian}). For this case, the bare $v_{sc}$ can be estimated as $v_{sc} \sim J^2_\chi \Phi_{ij}$ (the blue and red dot-dashed lines in Fig. \ref{fig:rgflow}(a)). 
  The RG $\beta$ functions for weak coupling $J_\chi$ and $v_{sc}$ near $g_{c1}$ read (see Section B1 of of Ref. \onlinecite{supp})
    \begin{align}
            j_\chi^\prime =\left( -\frac{\epsilon}{2} \right) j_{\chi}+\frac{1}{2} j_{\chi}^3,~
           v_{sc}^\prime=-\epsilon v_{sc} + j_{\chi}^{2}v_{sc} + 12 v_{sc}^3,
           \label{eq:RGgc1}
        \end{align} 
        where $\epsilon \equiv d-z $ and $j_\chi^\prime \equiv dj_\chi/dl$ with $dl \equiv -d \text{ln} \Lambda$ and $\Lambda$ being the energy running cutoff. In Eq. (\ref{eq:RGgc1}), a positive (negative) coefficient implies a relevant (irrelevant) term. 
        We find three non-trivial fixed points: 
$\left(j^{*^2}_{\chi}, ~ v^{*^2}_{sc} \right)
=\left( 0, ~ 0 \right),~~P:~
\left( \epsilon, ~ 0  \right)$ and 
$Q:~\left( 0, ~ \epsilon/12 \right)$. The resulting RG flow is shown in Fig. \ref{fig:rgflow}(a). 
           
Close to $g_{c2}$, the transition occurs deeply in the Kondo phase where the mean-field Kondo correlations $\chi_i$ are finite and the mean-field RVB $\Phi$ is negligibly small. The transition is therefore driven by the fluctuating $\hat{\tilde{\Phi}}$ field, and can be described by $S_{sc}$ near $g_{c2}$ in Eq. (\ref{eq:fulllagrangian}) with $v_{sc}\sim \chi^2 J_\Phi$ (the red dot-dashed line in Fig. \ref{fig:rgflow}(b)). In the weak coupling $J_\Phi$ and $v_{sc}$ regime, the RG equations read:
\begin{align}
	j_\Phi^\prime=-\frac{d}{2} j_{\Phi}+4j_{\Phi}^{3}+v_{sc}^{2}j_{\Phi}, 
           v_{sc}^\prime=\left(z-\frac{d}{2} \right) v_{sc}+v_{sc}^{3}~ .
           \label{eq:RGgc2}
\end{align} 
The RG flow for Eq. (\ref{eq:RGgc2}) is shown in Fig. \ref{fig:rgflow}(b). Except the Gaussian fixed point, one finds a non-trivial fixed point of Eq. (\ref{eq:RGgc2}) at $R:~(j^{*2}_{\Phi}, ~v_{sc}^{*2}) = (d/8,~0)$. 

We analyze below how the strange superconductivity emerges under RG. 
First, near $g_{c2}$, the RG flows of Eq. (\ref{eq:RGgc2}) reveal that both $v_{sc}$ and $j_\Phi$ grow to $(\infty,~\infty)$, corresponding to the (mean-field) coexisting Kondo-RVB superconducting phase. The critical fixed point $R$ involves the appearance/disapearance of the RVB correlation deep in the Kondo regime. This ``RVB breakdown" QCP corresponds to the ``hidden" AF QCP when superconductivity is suppressed (see the two dashed lines converged at the QCP in Fig. \ref{fig:rgflow}(d)). 

The RG flow of Eq. (\ref{eq:RGgc1}) near $g_{c1}$ seems to qualitatively capture the salient features of the phase diagrams of CeCoIn$_5$ and CeRhIn$_5$. Near $g_{c1}$, the bare couplings follow the relation: $v_{sc} \sim \Phi_{ij}J_{\chi}^2$. For a large mean-field value of $\Phi_{ij}$ (blue dot-dashed line), over a finite range in the parameter space, we find that $j_\chi$ and $v_{sc}$ flow to $(j_\chi,~v_{sc}) = (0,~\infty)$, a fixed point for superconductivity with Kondo breakdown. We expect that, without Kondo effect, the short-range RVB-spin liquid phase in this case becomes unstable against antiferromagnetism, likely leading to a coexstence of long-range AF and superconductivity [\onlinecite{foot-CeRhiN5-AF}]. Exprimental signatures of the Kondo breakdown-jump of the Fermi surface at critical pressure [\onlinecite{CeRhIn-FS-jump-shishido}]-as well as the the coexistence of the long-range AF order and superconductivity (illustrated in Fig. \ref{fig:rgflow}(c)) over a finite range in pressure have been observed in CeRhIn$_5$   [\onlinecite{Thompson-CeRhIn, Thompson-njp-2009}]. The Kondo breakdown transition is controlled by the vertical flow connecting the fixed point $P$ and the expected strong coupling fixed point (open circle) in Fig. \ref{fig:rgflow}(a), corresponding to $g^\prime$ in Fig. \ref{fig:rgflow}(c).
For smaller values of $\Phi_{ij}$ (the red dot-dashed line in Fig. \ref{fig:rgflow}(a)), however, a negligibly small coexisting AF-superconductivity phase in the parameter regime of  bare $J_\chi$ and $v_{sc}$ is found; it likely corresponds to the case of CeCoIn$_5$ at ambient pressure (as in Fig. \ref{fig:rgflow}(d)) where the superconducting phase does not coexist with AF order (except for the Q phase [\onlinecite{ThompsonPRL-2011}] and the field-induced AF order [\onlinecite{Shishido-PRL-Ce115}], which is out of our scope). The superconducting dome exhibits a hidden AF QCP inside, consistent with the experiments on CeCoIn$_5$  [\onlinecite{ThompsonPRL-2011, CeCoIn-Hall-Fisk, foot-Cddoping-SC-CeCoIn, Thompson-Cd-CeCoIn5}]. When the superconductivity is suppressed, the linear-in-field crossover scale seen in CeCoIn$_5$ near the QCP [\onlinecite{ThompsonPRL-2011}] can also be qualitatively accounted for within the RVB breakdown scenario via tuning $j_\Phi$ under RG at a finite Kondo correlation [\onlinecite{CHC-strangemetal}]. Note that the signatures of the proposed SL region in Fig. \ref{fig:rgflow}(c) and (d) are yet to be confirmed experimentally. A broad distribution in spinon-like excitation (instead of a sharp magnon-like excitation) spectrum [\onlinecite{Chung-spinon-stat}] in neutron scattering experiments [\onlinecite{Aeppli-neutronscatt}] would serve as a convincing evidence for the existence of SL region.  


{\it Conclusion.} 
We propose a microscopic mechanism for the strange superconductivity observed in the heavy fermion ``115" family (CeCoIn$_5$ and CeRhIn$_5$) based on the coexistence and competition between Kondo and short-ranged antiferromagnetic RVB correlations near an antiferromagnetic quantum critical point via the mean-field approximation and the field-theoretical renormalization group (RG) approaches to the Sp($N$) Kondo-Heisenberg model with valence fluctuations . 
The coexistence of these two phases underlie to superconductivity, while their competition gives rise to the 
antiferromagnetic quantum critical point and the quantum critical strange metal behaviors. Our mean-field theory describes comparatively well the superconducting phase of CeCoIn$_5$. We further construct an effective field theory beyond the mean-field level to reveal how superconductivity emerges from the strange metal state and its link to quantum criticality via the interplay of  coexistence and competition of the Kondo and RVB physics. Our RG analysis near the antiferromagnetic quantum critcal point can simultaneously describe the qualitative nature of the superconducting phases with and without coexisting antiferromagnetism for CeRhIn$_5$ and CeCoIn$_5$, respectively [\onlinecite{Thompson-CeRhIn}]. Our theory provides a critical step towards resolving the long standing issues on strange superconductivity in heavy fermion materials. Further experimental input is needed to clarified this issue.

{\it Acknowledgments.}
We thank P. Coleman, B. Maple, Y. Matsuda, F. Ronning, and J. Thompson for helpful discussions. This work is supported by the MOST grant No. 104-2112- M-009-004-MY3, the MOE-ATU program, the NCTS of Taiwan, R.O.C. (CHC). S. Kirchner acknowledges support by the National Science Foundation of China, Grant No. 11474250 and No. 11774307 and the National Key R\&D Program of the MOST of China, Grant No. 2016YFA0300202 and No. 2017YFA0303104.


\begin{thebibliography}{99}
\bibitem{footnote-Tk-1-ion}
Note that $T^\ast_{\text{onset}}$ here is different from the single-ion Kondo temperature $T_K^{1-ion}\sim 1.5$ K of a dilute system [\onlinecite{La-doped-CeCoIn}].
\bibitem{foot-T-onset}
Here, we expect $T^\ast_{\text{onset}}$ will be enhanced when the crystal electric field effect is included [\onlinecite{Feng-arpes-CeCoIn, Coleman-book, Satoh-Tk, HewsonBook}]. Experimentally, the onset temperature $T_{\text{onset}}^\ast$ of the Kondo hybridization has been found to start at more than 200K [\onlinecite{Feng-arpes-CeCoIn}] in the presence of CEF effect,
\bibitem{foot-nf-p9}For a $0.05\%$ doping of Cd [\onlinecite{Thompson-Cd-CeCoIn5}], we expect the average number of unpaired electrons per the Ce atom increases by 0.15 since some In$^{1+}$ ions in the CeIn$_3$ layer are replaced with Cd$^{2+}$. The valence fluctuation (0.8 $\sim$ 0.9) leads to a nearly half-filled in the local $f$-band: The average occupation number $n_f$ per site now becomes $n_f = (1+ 0.15) \times 0.8 = 0.92$ to $n_f = (1+ 0.15) \times 0.9 = 1.03$. Hence, taking the Lagrange multiplier $\lambda \rightarrow \infty$ in our RG calculation is reasonable.
\bibitem{foot-quasi-2d} The quasi-2d nature of the system is supported by the large spin anisotropy in CeCoIn$_5$, which has been experimentally confirmed from the spin susceptibility measurement [\onlinecite{Thompson-CeCoIn-SC}].
\bibitem{foot-CeRhiN5-AF}To fully confirm theoretically the coexisting AF and superconductivity for CeRhIn$_5$, the long-range AF order must be included in our Hamiltonian, which will be addressed separately elsewhere.
\bibitem{foot-Cddoping-SC-CeCoIn}Upon Cd doping, the coexistence between AF and superconductivity in CeCoIn$_5$ has been observed via neutron scattering. See, for example Ref. \onlinecite{Thompson-Cd-CeCoIn5}.
\end{thebibliography}
    \end{document}